\documentclass[12pt]{article}

\usepackage{amssymb,amsmath}
\usepackage{fullpage}
\usepackage{graphicx}
\usepackage{amsmath}		
\usepackage[margin=1.0in]{geometry}
\usepackage{setspace}
\usepackage{color}
\usepackage{fancyhdr}
\usepackage{collcell}
\usepackage{datatool}
\usepackage{environ}
\usepackage{latexsym}
\usepackage{amssymb}
\usepackage{epsfig,amsmath,graphics}
\usepackage{epstopdf}
\usepackage{verbatim}
\usepackage{wasysym}
\usepackage{feynmp-auto}
\usepackage{authblk}
\usepackage{xcolor}
\usepackage{enumitem}
\usepackage[utf8]{inputenc}
\usepackage{slashed}
\usepackage{cite}
\usepackage[toc,page]{appendix}
\usepackage[normalem]{ulem}

\usepackage{empheq}

    \newlength\fsep
    \newsavebox\widebox

\newcommand{\Rt}{\phantom{!}^{^{(3)}}\!\! R}

\definecolor{deepgreen}{rgb}{0.2,0.8,0.3}

\definecolor{deepblue}{rgb}{0.2,0.2,0.8}

\definecolor{deepred}{rgb}{0.8,0.2,0.2}

\usepackage[skip=0pt]{caption}

\begin{document}
\title{The Classical Equations of Motion of Quantized Gauge Theories, Part I: General Relativity}
\author[1]{David E.~Kaplan}
\author[2]{Tom Melia}
\author[1]{Surjeet Rajendran}
\affil[1]{\small Department of Physics \& Astronomy, The Johns Hopkins University, Baltimore, MD  21218, USA}
\affil[2]{\small Kavli IPMU (WPI), UTIAS, The University of Tokyo, Kashiwa, Chiba 277-8583, Japan}

\date{\today}

\maketitle

\begin{abstract}
In this and a companion paper, we show that quantum field theories with gauge symmetries permit a broader class of classical dynamics than typically assumed.  In this article, we show that the dynamics extracted from the path integral or Hamiltonian formulation of general relativity allows for classical states that do not satisfy the full set of Einstein's equations. This amounts to loosening the Hamiltonian and momentum constraints that are imposed on the initial state. Nevertheless, the quantum theory permits gauge invariant time evolution of these states. The time evolution of these states is such that at the classical level the full set of Einstein's equations would appear to hold, with the physical effects of these states being attributable to an auxiliary, covariantly conserved energy-momentum tensor with no internal degrees of freedom.  We derive the generalized Einstein equations for these states and show  that a homogeneous and isotropic initial background state contributes to expansion identical to cold dark matter. The inhomogeneous components of this state could source curvature perturbations that grow linearly at linear order. This auxiliary contribution to Einstein's equations could have either sign and thus provide a trivial way to violate the null energy condition, enabling novel gravitational dynamics such as cosmic bounces and wormholes.

\end{abstract}

\tableofcontents


\section{Introduction}
Why do we observe nature obeying certain classical equations of motion? At the level of classical mechanics, these equations are axioms and our belief in them is the result of experimental validation. But, the fundamental laws of nature are quantum mechanical and classical physics is merely a limit of the quantum theory. In this context, the classical equations of motion are not axiomatic. Instead, they are identities that relate the expectation values of various quantum mechanical operators. These identities are automatically enforced whenever the quantum state evolves per the sole axiomatic equation of quantum mechanics - Schr\"{o}dinger's equation. This equation governs the time evolution of all quantum states, including that of macroscopic, classical systems. In particle mechanics, these identities are proven by Ehrenfest's theorem. For field theories, similar identities can be obtained through the Schwinger-Dyson procedure. The purpose of this paper (and its companion) is to highlight an important subtlety that arises when this procedure is applied to quantum field theories that possess gauge symmetries and explore the consequences of this subtlety. 

The equations of motion of a classical field theory are obtained by taking a classical action, varying it and setting the variation to zero. In field theories with a gauge symmetry, some of the degrees of freedom are not physical - they can be gauged away.  As a result, there are fewer second order dynamical equations of motion than the number of fields used to represent the theory in the classical action - these equations govern the evolution of the physical degrees of freedom of the theory. Naive variations of the action along the gauge degrees of freedom yield additional equations - but, since there are no associated physical degrees of freedom corresponding to these, the resultant equations of motion take the form of constraint equations on the physical degrees of freedom. There are thus more equations than available degrees of freedom and one might worry that this system of equations cannot be consistently solved. However, it can be shown that if the dynamical equations are given initial conditions that obey the constraints, the evolution equations, in concert with the conservation equations associated with the gauge symmetry, will ensure that the time evolution continues to automatically obey the constraints. 

What if the evolution equations were given initial conditions that did not obey the constraints? The second order equations for the physical degrees of freedom can still be solved and one would obtain unambiguous time evolution for these states. In classical mechanics we would reject such states simply because of their failure to obey the classical constraints. But, classical mechanics is a limit of quantum mechanics and thus the correct question to ask is if the classical equations of motion emerge as identities of the quantum dynamics described by Schr\"{o}dinger's equation. Since Schr\"{o}dinger's equation is a dynamical equation, it can enforce the dynamical equations of the classical theory but not the constraint equations. Do the constraint equations follow from the quantum theory? In conventional quantization schemes, these constraint equations are imposed as additional restrictions on the quantum states of the theory. These restrictions are motivated by a desire to reproduce the observed classical laws of nature and justified by the belief that these restrictions are necessary to obtain gauge invariant physics. 

But, the only logical restriction that needs to be imposed on the quantum states of the theory are that they lead to well defined, gauge invariant physics. The quantum theory, being the deeper underlying theory, can deviate away from classical physics. With this relaxed but logically consistent criterion, we show that there are a broader class of quantum states that can be consistently evolved in gauge theories than conventionally assumed.  These states do not obey the constraint equations, but they are nevertheless well defined and give rise to gauge invariant physics. The time evolution of these states will violate the constraint equations - but, as we will see, the gauge symmetry and the associated conservation equations restrict these variations to a simple class of deviations. These deviations have observable consequences and it is thus a matter of experiment to see if the quantum state of our universe belongs to the set of quantum states that are restricted to obey these  constraint equations or if the state belongs to this broader class that violates these constraints.

In this paper, we focus on general relativity while the companion paper focuses on electromagnetism. The rest of this paper is organized as follows. In Section \ref{sec:Schwinger}, we review the derivation of the classical equations of motion using the Schwinger-Dyson procedure and highlight the subtleties that arise in applying this procedure to gauge theories. Then, proceeding with the quantization of general relativity, in Section \ref{sec:mini}, we apply this mathematical framework to a toy model of gravity, namely minisuperspace, which describes a homogeneous and isotropic cosmology. With the understanding from this section, we derive the quantum dynamics of gravity in Section \ref{sec:MGR}.  In the final section, we conclude and discuss a broader set of phenomenological questions for such theories. We include two appendices for the sake of completeness. In Appendix \ref{sec:Mani}, we describe the mathematical framework necessary to describe general relativity as a quantum field theory on a manifold and show how the theory can be described in a gauge invariant manner. In Appendix \ref{sec:Coherent}, we describe the  coherent state basis in minisuperspace.   Throughout we use Greek letters for four-vector indices, Latin letters for three-vector (spatial) indicies, and overdots for coordinate time derivatives.  We use standard notation for normal partial derivatives $_{,\mu} \equiv \partial/\partial x^\mu$.


\section{The Classical Limit}
\label{sec:Schwinger}
In this section, we review the Schwinger-Dyson procedure that shows how the classical equations of motion of a field theory are automatically obeyed by expectation values of various quantum field operators whenever the time evolution of the state obeys  Schr\"{o}dinger's equation. Consider, for example,  the theory of a scalar field $\phi$. In the quantum theory, the field and its conjugate momentum are promoted to operators $\hat{\phi}$ and $\hat{\pi}$ respectively. A Hamiltonian $\hat{H}$ is constructed from these operators. Quantum states, constructed for example as a Fock space using the operators $\hat{\phi}$ and $\hat{\pi}$, evolve as per  Schr\"{o}dinger's equation. It is then straightforward to see that the following equations hold: \footnote{We are loose with notation here.  The final terms should be interpreted as functional derivatives of the classical Hamiltonian with respect to fields $\phi$ and $\pi$ and then replacing all fields with operators, keeping ordering fixed.}
\begin{eqnarray}\label{eq:HamEOM1}
      \partial_t \langle \hat{\phi} \rangle = i \left\langle \left[ \hat{H} , \hat{\phi} \right] \right\rangle =  \left\langle \frac{\partial\hat{H}}{\partial\hat{\pi}} \right\rangle \\ \label{eq:HamEOM2}
     \partial_t \langle \hat{\pi} \rangle = i \left\langle \left[ \hat{H} , \hat{\pi} \right] \right\rangle = -  \left\langle \frac{\partial\hat{H}}{\partial\hat{\phi}} \right\rangle
\end{eqnarray}
These are Hamilton's equations of motion in expectation value. This can also be shown using the path integral approach. In the path integral approach, the time evolution operator $T\left(t_2;t_1\right)$ that describes the time evolution of a quantum state from time $t_1$ to $t_2$ is defined in terms of its transition matrix elements on field basis states $|\phi\rangle$. This transition matrix element is given by: 

\begin{equation}
    \langle \phi_f | T\left(t_2;t_1\right) |\phi_i\rangle = \int_{\phi\left(t_1\right) = \phi_i}^{\phi\left(t_2\right) = \phi_f} D\phi \,  e^{i S[\phi]}
    \label{eq:PI}
\end{equation}
where $S =\int {\cal L}d^4 x$ is the classical action. Redefine the field variables in the integral by a shift $\phi \rightarrow \phi + \delta\phi$ where $\delta\phi$ is taken to vanish at the  boundaries of integration. The variable shift should not change the physics of the quantum states. By setting this variation to zero, we get the following identity:
\begin{equation}\label{eq: Schwinger-Dyson}
    \int {\cal D}\phi e^{i S\left[\phi\right]} = \int {\cal D}\phi e^{i S\left[\phi\right] + i \delta S\left[\phi + \delta\phi\right]} \implies \int {\cal D}\phi \left(1 + i\frac{\delta S}{\delta\phi} \delta\phi \right) e^{i S\left[\phi\right]} = 0
\end{equation}
Thus, for an arbitrary (but infinitesimal) variation $\delta\phi$, we have 
\begin{equation}\label{eq:EulEOM}
    \left\langle \frac{\delta S}{\delta\phi} \right\rangle = \left\langle \frac{\delta{\cal L}}{\delta\phi} - \partial_\mu \frac{\delta{\cal L}}{ \delta\left(\partial_\mu  \phi\right)} \right\rangle = 0
\end{equation}
In other words, the classical equations of motion, derived by demanding that the classical action is stationary with respect to variations of any field in the action, is satisfied at the expectation value level. 

Observe the following subtlety in applying this procedure to theories with gauge symmetries. In these, the number of physical degrees of freedom are smaller than the number of fields used to represent the theory in the classical action.  In the quantum theory, these superfluous or gauge degrees of freedom are identified by the fact that these degrees of freedom do not posses corresponding conjugate momenta.  Since these degrees of freedom are not physical, they need to be fixed through some procedure in order to define a Hamiltonian. But, once the degree of freedom is fixed, what does it mean to vary the Hamiltonian with respect to this fixed degree of freedom as per \eqref{eq:HamEOM1}? If the quantum theory is defined using the path integral, the extraction of the transition matrix elements  \eqref{eq:PI} requires the gauge degrees of freedom to be factorized. This factorization effectively enforces a gauge choice on the path integral. Thus the action that appears in the path integral that defines the transition matrix elements \eqref{eq:PI} is not the original classical action $S$ but instead a suitably gauge fixed action $S_{gf}$. Since $S_{gf} \neq S$, the identities implied by changing the variables of integration in the path integral are different from the conventional classical equations obtained by varying $S$ (see discussion in \cite{Burns:2022fzs}).

Note the key distinction between the above quantum method of deriving the equations of motion and the classical approach. In the classical theory, the classical action is first varied and the equation of motion is obtained by setting these variations to zero. A gauge choice is subsequently made to solve these equations of motion. In the quantum theory, the definition of the Hamiltonian or the path integral requires a gauge choice -- without such a choice, either the Hamiltonian has ill-defined operators or the gauge degeneracies have not been factorized from the path integral. This changes the derived classical equations. In the following, we describe how these issues impact the quantization of general relativity and the extraction of the classical limit of this quantum theory.

\section{Minisuperspace}
\label{sec:mini}

We start the discussion with a toy (but illustrative) model of cosmology known as minisuperspace.  It is general relativity in four dimensions and it describes homogeneous and isotropic space-times.  Namely, it is generated from a metric
\begin{equation}
    g_{\mu \nu} \rightarrow ds^2 = - N\left(t\right)^2 dt^2 + a\left(t\right)^2 \left(dx^2 + dy^2 + dz^2\right)
    \label{Eqn:GRMetric}
\end{equation}
The spatial degrees of freedom are frozen and the theory describes the time evolution of the scale factor $a$ in concert with other homogeneuous and isotropic forms of matter. The action is the  standard Einstein-Hilbert term plus matter
\begin{equation}
    S_{ms}=\int dt \sqrt{-g}\left({M_{pl}^2}  R + {\cal L}_{matter} \right)
\end{equation}
where $M_{pl}=(8\pi G_N)^{-1/2}$ is the reduced Planck mass, and $R$ and $g$ are the Ricci scalar and the metric determinant respectively. The matter fields are position independent and the volume integral (an overall volume factor) has been removed. 

Finding the paths that extremize the action leads to the standard Friedmann equations. For example, for a scalar field $\phi$ with matter action  $\mathcal{L}_{matter} = -\frac{1}{2}g^{\mu \nu} \partial_{\mu}\phi \partial_{\nu}\phi -  V(\phi)$, and the metric \eqref{Eqn:GRMetric}, the full action is\footnote{We have tacitly integrated by parts to remove terms with $\ddot a$.  This is equivalent to including a Gibbons-Hawking-York term needed to produce the correct equations of motion for manifolds with boundaries \cite{Gibbons:1976ue,York:1972sj}.}
\begin{equation}\label{eq:Sms}
    S_{ms} = \int dt \left(- 6 M_{pl}^2\frac{a(t) \dot{a}(t)^2}{N(t)} + \frac{a(t)^3 \dot{\phi}(t)^2}{2 N(t)} - N(t) a(t)^3 V(\phi)\right)
\end{equation}
leading to classical equations of motion (dropping the $t$ arguments of $N$ and $a$ for clarity)
\begin{eqnarray}
    \frac{\delta S_{ms}}{\delta N} &=&  a^3 \left( 6 M_{pl}^2\frac{ \dot{a}^2}{N^2 a^2} - \frac{ \dot{\phi}^2}{2 N^2} - V(\phi)\right) = 0 \label{eq:Fried1}\\ \nonumber \\ 
    \frac{\delta S_{ms}}{\delta a} &=& 3 N a^2 \left(  4 M_{pl}^2\frac{ \ddot{a}}{N^2 a} + 2M_{pl}^2 \frac{\dot{a}^2}{N^2 a^2} - 4 M_{pl}^2 \frac{\dot{a}\dot{N}}{N^3 a} +  \frac{ \dot{\phi}^2}{2 N^2} -  V(\phi)\right) = 0 \label{eq:Fried2}\\ \nonumber
    \\
    \frac{\delta S_{ms}}{\delta\phi} &=& - \frac{a^3}{N}\left( \ddot{\phi} + 3\frac{\dot{a}}{a} \dot{\phi} - \frac{\dot{N}  \dot{\phi}}{N} + N^2 \frac{\partial V(\phi)}{\partial \phi}\right) = 0
\end{eqnarray}
Upon fixing a time coordinate by choosing, for example, $N(t) = 1$, the above reduce to Friedmann's first and second equations and the equation of motion for the scalar.

While the above is the standard procedure, we have been a bit cavalier.  If we take the action $S_{ms}$ and redefine the path for this integral $N(t) \rightarrow N(t) + \delta N(t)$ (with vanishing $\delta N$ at the boundaries), we would require the shift in the action $S_{ms}\left[N,a,\phi\right] \rightarrow S_{ms}\left[N+\delta N,a,\phi\right]$ to be trivial.  However, it is already true that $S_{ms}\left[N+\delta N,a,\phi\right] = S_{ms}\left[N,a,\phi\right]$ for {\it any path}.  This is because a change in path of $N$ can be compensated for by defining a new time coordinate integration variable $dt' = (1 + \frac{\delta N}{N})dt$.  This suggests that in the quantum theory, the path integral does not guarantee this first equation in the classical limit.  We will see how this bears out below.

\subsection{Canonical Quantization}

Now we want a quantum version of this theory.  We see that $\dot{N}(t)$ does not appear in the action and thus is not dynamical.  It also plays a central role in the reparameterization invariance of the time coordinate, (namely, $dt\rightarrow f(t)dt$ and $N(t)\rightarrow N(t)/f(t)$), and thus is something like a gauge degree of freedom.  To canonically quantize, let's first look at the classical Hamiltonian.  On the metric side, only $a(t)$ has a conjugate momentum, $\pi = \delta{\cal L}/\delta \dot{a} = - 12 M_{pl}^2 a \dot{a}/N$, since $\delta{\cal L}/\delta \dot{N} = 0$.  The scalar's conjugate variable is $\pi_{\phi} = \delta{\cal L}/\delta \dot{\phi} = (a^3/N) \dot{\phi}$.
The resulting Hamiltonian
\begin{equation}\label{eq: mini-H}
    H = \left[\pi \dot{a} + \pi_\phi \dot{\phi} -{\cal L} \right]_{\dot{a} = \cdots, \dot{\phi} = \cdots} = - \frac{N}{24 M_{pl}^2 a} \pi^2 + \frac{N}{2 a^3} \pi_\phi^2 + N a^3 V(\phi)
\end{equation}
is interestingly  $-N (\delta S_{ms}/\delta N) \equiv N \tilde{H}$, or $N$ times the function that vanishes by the first Friedmann equation \eqref{eq:Fried1} when replacing $\dot{a}$ and $\dot{\phi}$ by functions of their conjugate momenta. 

In the quantum theory, there is now the issue of what to do with $N$, which appears as an auxiliary field that naively commutes with all operators.  The natural thing to do is fix a gauge, noting that the Schr\"{o}dinger equation with the above Hamiltonian would be of the form:
\begin{equation}
    i\partial_t |\psi\rangle = N(t) \hat{\tilde{H}} |\psi\rangle
\end{equation}
and is covariant with respect to time reparameterizations.  Thus, to define the quantum theory, a sensible choice is to pick a time coordinate by way of fixing $N(t)$.

Now, once we fix $N$, we can write a quantum Hamiltonian operator by interpreting $a, \pi, \phi, \pi_\phi$ as operators with standard commutation relations.  There are remaining issues, such as defining inverse field operators and operator ordering, which we will address below when we derive the path integral.

Our main point is that once $N$ is fixed, we only get the following equations in the gravitational sector:
\begin{eqnarray}
    \partial_t \langle \hat{a} \rangle = -i \langle \left[ \hat{\Tilde{H}},\hat{a}\right] \rangle \\
    \partial_t \langle \hat{\pi} \rangle = -i \langle \left[ \hat{\Tilde{H}},\hat{\pi}\right] \rangle 
    \label{eqn:QuantumEquations}
\end{eqnarray}
 When appropriately defining the commutators, inverse operators, and operator ordering, the first equation reproduces the definition of the conjugate field $\pi$, while the second equation reproduces the second Friedmann equation.  The first Friedmann equation would require $\langle \hat{\Tilde{H}}\rangle = 0$, but is not necessarily a consequence of this quantum theory. In the classical theory, this equation is produced by varying $N$ - but in the quantum theory, to write down the Schr\"{o}dinger equation we had to pick a time parameterization and thus fix $N$. Once $N$ has been fixed, there aren't additional equations that can be derived from varying it (see discussion in section \ref{sec:Schwinger}). Of course, the theory can be defined with any choice of $N$ - but this simply changes the definition of time and does not restrict the physical states of the theory.  
 
 While not guaranteed by the above theory, what is guaranteed is
\begin{equation}
    \partial_t \langle \hat{\Tilde{H}}\rangle = i \langle \left[ \hat{\Tilde{H}},\hat{\Tilde{H}}\right]\rangle = 0
\end{equation}
which means if the initial state is chosen to satisfy the first Friedmann equation (in expectation value), then it satisfies it at all times.  This is equivalent to the classical equations, where the time derivative of the first Friedmann equation is  linearly dependent of the second Friedmann equation and matter equations of motion.  Thus, one can impose $\langle \hat{\Tilde{H}}\rangle = 0$ on the initial state and it will be satisfied at all times.  Note, classically this is just fixing the initial boundary condition for $\dot{a}(t)$.

What is instead often imposed in the literature is a second order constraint -- namely that the Hilbert space is restricted to the states $|\psi\rangle$ such that $\hat{{H}} |\psi\rangle = 0$.  This is known as the Wheeler-DeWitt equation \cite{DeWitt:1967yk}, and while at the equation level seems to reproduce classical physics, it in fact generates a theory without time evolution -- the Schr\"{o}dinger equation becomes trivial and the quantum state is static. It has been suggested that time could emerge from this ``timeless'' quantum state by the entanglement of one degree of freedom with the other degrees of freedom so that the relative evolution between these states would appear as time evolution. This suggestion has never been actually implemented and it has largely been ruled out in general terms \cite{Unruh:1989db}. Moreover, while the idea of time being connected to the relative evolution of quantum states has merit, the restriction imposed by the Wheeler-DeWitt equation that the quantum state remain static implies that the only permitted states of the theory are those where the quantum state is a macroscopic superposition over the entire history of the cosmos \cite{Page:1983uc}.

\subsection{Coherent States and a Path Integral Description}
In quantum mechanics (or quantum field theory), a path integral represents the time evolution operator, typically in the general coordinate (or field) basis.  It can be derived from a Hamiltonian in a fixed gauge, or can be gauge-fixed in situ.  Below, we will instead derive the path integral in the coherent state basis, which makes the classical-like evolution of these states even more manifest.

For minisuperspace, we can start by taking the quantum version of the Hamiltonian \eqref{eq: mini-H} with $N=1$, and derive the path integral.  At this point we are required to define the operators that comprise the kinetic terms. To do so, we will define a basis of states which are a set of coherent states: $|\pi_{cl} a_{cl}\rangle$.  One can define these states using standard tricks from the quantum harmonic oscillator.  First, define a `ground state' $|0\rangle$ as a state annihilated by the operator $\hat{b} \equiv (\hat{a} + i \hat{\pi})/\sqrt{2}$, which obeys  $\left[ \hat{b},\hat{b}^\dagger \right] = 1$.  One can now build coherent states in the standard way by finding eigenstates of $\hat{b}$:
\begin{eqnarray*}
    & |\beta\rangle = e^{-|\beta|^2/2} e^{\beta \hat{b}^\dagger} |0\rangle \\
    & \hat{b}|\beta\rangle = \beta |\beta\rangle
\end{eqnarray*}
where $\beta$ is a c-number.  One can easily show that $\langle\beta | \hat{a} | \beta\rangle = (\beta + \beta^*)/\sqrt{2}$ and $\langle\beta | \hat{\pi} | \beta\rangle = -i (\beta - \beta^*)/\sqrt{2}$.  Thus, we define the real and imaginary parts of $\beta$ as $\beta_r \equiv a_{cl}/\sqrt{2}$ and $\beta_i \equiv \pi_{cl}/\sqrt{2}$, respectively.

We now define a Hamiltonian that will generate on average classical time evolution by defining operators in terms of coherent states, which will ensure such states evolve dominantly in a classical way. It is known that the coherent state basis is over-complete, and that an operator can be fully defined in this basis by its diagonal elements (a brief proof is in Appendix \ref{sec: CoStOps}).  Thus, taking the gauge-fixed version of the classical Hamiltonian \eqref{eq: mini-H}, an initial proposal for the quantum Hamiltonian of mini-superspace is
\begin{equation}\label{eq:QH-ms}
    \hat{H} = - \frac{1}{24 M_{pl}^2} \widehat{\left[ \frac{\pi^2}{a}\right]} + \frac{1}{2} \hat{\pi}_\phi^2 \widehat{\left[\frac{1}{a^3}\right]} + \widehat{\left[a^3\right]} V(\hat{\phi})
\end{equation}
where the operators in square brackets are defined by their diagonal matrix elements in the $|\pi_{cl} a_{cl}\rangle$ basis, {\it e.g.},
\begin{equation}
    \langle \pi_{cl} a_{cl} | \widehat{\left[\frac{\pi^2}{a}\right]} | \pi_{cl} a_{cl} \rangle = \frac{\pi_{cl}^2}{a_{cl}}
\end{equation}
and they commute with matter operators $\hat{\phi}$ and $\hat{\pi}_\phi$.  

It turns out that for simply defined coherent states (which in the field basis, do not vanish at $a=0$ and have exponentially small support in the $a<0$ region), the matter kinetic term operator is ill-defined.  There are two ways to address this.  The first is to use a more sophisticated coherent state basis, one whose states vanish as you approach $a\rightarrow 0$.  Such coherent states have been build for the coherent states of the hydrogen atom, namely addressing the radial coordinate \cite{Brown:1973xx,Klauder:1995yr}, and one could build similar states here.  A perturbative approach is to shift the scale factor operator by a fixed value $\hat{a}\rightarrow a_0 + \hat{\Delta a}$.  This amounts to a change of basis of coherent states, as the operator $\hat{\delta a}$ still has the standard commutator with $\hat{\pi}$.  Now, operators can be Taylor expanded, {\it e.g.},
\begin{equation}
    \widehat{\left[ \frac{\pi^2}{a_0 + \Delta a} \right]} \equiv \frac{1}{a_0}\sum_n (-1)^n \frac{\widehat{\left[\pi^2 \Delta a^n\right]}}{a_0^{n}}
\end{equation}
where each term in the sum is again defined by its diagonal matrix elements.  These operators are now perturbatively  well-defined (both on and off diagonal matrix elements).  Once expanded in fact, all terms in the power series have positive powers of $\hat{\pi}$ and $\Delta \hat{a}$.  Thus, these bracketed operators are equivalent to the normal ordered operator, {\it e.g.}, $\widehat{\left[\pi^2 \Delta a^n  \right]} \equiv \; :\! \hat{\pi}^2 \Delta\hat{a}^n\! :$ .  While these coherent states are not eigenstates of the Hamiltonian, they can be shown to evolve classically up to small wave-function spreading if the initial spread is small (which is true if $a_0\gg \Delta a_{cl} \gg 1$ with the above normalization).  This Hamiltonian and coherent state basis is useful when the initial expectation value $\langle\Delta \hat{a}\rangle = a_{cl} - a_0 \equiv \Delta a_{cl} \ll a_0$.  As the expectation value evolves outside the domain of validity of the Taylor expansion, one can match the state onto a new basis expanded around a different value of $a_0$.

Now we are poised to build the path integral using the coherent-state basis. Using the completeness relation for coherent states (for example, for the metric fields):
\begin{equation}
    \frac{1}{\pi}\int \left[ d\pi\right] \left[ da \right] | \pi a \rangle \langle \pi a | = 1
\end{equation}
one can break a transition matrix element into infinitesimal time steps $\delta t$ by inserting a complete set of coherent states, as we do in Appendix \ref{sec:Coherent}.  The resulting path integral is
\begin{equation}\label{eq:msPI}
    \langle \pi_f a_f \pi_{\phi_f} \phi_f| e^{-i \hat{H} t} | \pi_i a_i \pi_{\phi_i} \phi_i\rangle = \int_{\pi_i,a_i,\pi_{\phi_i}, \phi_i}^{\pi_f,a_f,\pi_{\phi_f}, \phi_f} \, {\cal D}\pi \, {\cal D}a \,{\cal D}\pi_\phi \, {\cal D}\phi \, e^{i \int dt \left[\frac{1}{2}(\dot{a}\pi - \dot{\pi}a + \dot{\phi}\pi_\phi - \dot{\pi}_\phi\phi) - H\left[\pi,a,\pi_\phi,\phi\right]\right]}
\end{equation}
where the boundaries of the path integral are labeled by the expectation values of $\hat{\pi}$ and $\hat{a}$ in the initial and final states.  For fixed initial values, $a_i$ and $\pi_i$, this path integral can be taken to be a wave function of $a_f$ and $\pi_f$ in coherent state space.  It is easy to show that for a given initial state, this wave function is sharply peaked around the classical predictions for $a_f$ and $\pi_f$ dictated by the second Friedmann equation.  If one wants the system to satisfy the first Friedman equation, one can choose an initial value of $\pi_i$ to satisfy this constraint, after which the $\pi_f$ peak value will also satisfy this constraint.

We can now write the resulting classical equations of motion for this theory, first by solving the equations for $\pi$ and $\pi_\phi$.  What is new is that the first Friedmann equation is not required to be satisfied (but its time derivative is).   We will define the amount of violation by the constant $\mathbb{H}_0$, allowing us to write the equations:
\begin{eqnarray}
   6 M_{pl}^2 a \dot{a}^2 = a^3(\frac{1}{2}\dot{\phi}^2 + V(\phi)) + \mathbb{H}_0 \label{eq:FirstFried}\\
    -2M_{pl}^2(2 a \ddot{a} + \dot{a}^2) =  a^2 (\frac{1}{2}\dot{\phi}^2 - V(\phi)) \\
    a^3(\ddot{\phi} + 3\frac{\dot{a}}{a}\dot{\phi} + \partial_\phi V(\phi)) = 0
    \label{eqn:miniequations}
\end{eqnarray}
What we see (after scaling out by the appropriate powers of $a$) is that the non-compliance with the first Friedmann equation or Hamiltonian constraint \eqref{eq:FirstFried} leads to a more general initial condition for $\dot{a}$, which could be interpreted as the existence of a new but non-substantive source of energy density that has zero pressure.  In other words, it contributes to the equations as a component of dust or a dark matter would.  Interestingly, the violation of \eqref{eq:FirstFried}, parameterized by $\mathbb{H}_0$ could take either sign, and thus could contribute as a substance with  negative energy density.  There are of course no additional fields here, but an quantum state of field configurations of the non-dynamical gravitational fields. 
Note, this and similar effects has also been seen looking at constrained gravitational instantons \cite{Cotler:2020lxj},  in Horava-Lifschitz gravity \cite{Horava:2009uw,Mukohyama:2009mz}, and in Einstein-aether theories \cite{Jacobson:2015mra,Speranza:2015sta}.



\section{General Relativity}\label{sec:MGR}

We will now take what we learned from the minisuperspace  example and look at the classical limits from a quantized field theory with a symmetric (metric) tensor, $g_{\mu\nu}$, coupled to itself and matter in a general coordinate invariant way.

The classical action that generates general relativity is of course the Einstein-Hilbert action
\begin{equation}\label{eq:E-H}
    S =\int d^4x  \sqrt{-g} \left({M_{pl}^2}R + {\cal L}_{matter} \right) + S_{GHY}
\end{equation}
 where the matter Lagrangian can contain independent fields as well as the metric, its inverse, and covariant derivatives associated with the metric.  We have also included the final Gibbons-Hawking-York term explicitly to remove boundary terms.  Minimization of this action with respect to the path of the metric tensor produces Einstein's equations:
\begin{equation}
    \frac{\delta S}{\delta g_{\mu\nu}} = \sqrt{-g}\left( M_{pl}^2  G^{\mu\nu} - T^{\mu\nu}\right) = 0
\end{equation}
where from the second term in \eqref{eq:E-H}, $S_{m} =  \int\sqrt{-g}{\cal L}_{matter}$:
\begin{equation}
    T^{\mu\nu} = - \frac{1}{\sqrt{-g}}\frac{\delta S_{m}}{\delta g_{\mu\nu}}
\end{equation}
These plus the matter equations of motion determine the classical dynamics as a function of initial conditions. As in the classical version of minisuperspace, some of these equations are in fact constraints on the initial conditions.

To canonically quantize, one must find a suitable Hamiltonian operator.  Like the previous examples of minisuperspace and QED, there are redundancies in this theory in the form of reparameterization invariance.  The coordinates can be redefined in terms of four independent functions, $x^\mu \rightarrow \xi^\mu (x)$.  This imposed invariance also results in some components of $g_{\mu\nu}$ being non-dynamical or lacking standard kinetic terms.  Thus we have a problem when attempting to define conjugate momenta for components of $g_{\mu\nu}$, namely that some vanish.  Define $L_{grav} \equiv \int d^3x (\sqrt{-g} M_{pl}^2 R + L_{boundary})$, where the last term is to remove higher derive/boundary terms.  As the matter part of the action does not contain time derivatives of metrics, the conjugate momenta for the metric components are naturally defined as:
\begin{align*}
    \pi^{ij}  &\equiv \frac{\delta L_{grav}}{\delta g_{ij,0}} \\ 
    \pi^{i}  &\equiv \frac{\delta L_{grav}}{\delta g_{0i,0}} = 0 \\ 
    \pi  &\equiv \frac{\delta L_{grav}}{\delta g_{00,0}} = 0
\end{align*}

There is now a clear choice to make -- effectively a gauge choice before we define the Hamiltonian operator.  By choosing $g_{0\mu} = - \delta^0_\mu$, we both remove the fields (traditionally the `lapse function' and `shift vector') without conjugates and get the benefit of choosing a time coordinate and time slices on which to identify initial states.

The gravitational part of the Lagrangian density in this gauge (now setting $M_{pl}=1$) can be shown to be

\begin{equation}
   {\cal L} =   \sqrt{\gamma}\left(\frac{1}{4} (\gamma^{ik}\gamma^{j\ell} - \gamma^{ij}\gamma^{k\ell})\gamma_{ij,0}\gamma_{k\ell,0} \, + \!\!\Rt \right) 
\end{equation}
where $\!\!\Rt$ is the three-dimensional Ricci scalar associated with the spatial metric $g_{ij}\equiv\gamma_{ij}$, $\gamma$ is the determinant of the spatial metric, and spatial indices are raised and lowered with this metric.

In this gauge we see that from the Lagrangian standpoint, the equations of motion are reduced to the spatial part of Einstein's equations, and thus these will be the only ones guaranteed in expectation value in the quantum theory.

We can now build a Hamiltonian operator for the quantum theory in terms of the conjugate momentum, which classically is
\begin{equation}
    \pi^{ij} = \frac{1}{2}\sqrt{\gamma}(\gamma^{ki}\gamma^{\ell j} - \gamma^{ij}\gamma^{k\ell} ) \gamma_{k\ell , 0}
\end{equation}
Solving for the time derivative of the metric
\begin{equation}
    \gamma_{ij,0} = \frac{2}{\sqrt{\gamma}} ( \gamma_{ia} \gamma_{jb} - \frac{1}{2} \gamma_{ij}\gamma_{ab})\pi^{ab}
\end{equation}
allows us to construct the classical Hamiltonian density
\begin{equation}
    {\cal H} = \frac{1}{\sqrt{\gamma}}(\pi_{ij}\pi^{ij} - \frac{1}{2}\pi^2) - \sqrt{\gamma} \Rt
\end{equation}
where $\pi\equiv \gamma_{ij}\pi^{ij}$. Now we would like to use the classical intuition to produce a Hamiltonian operator in a quantum theory.  Again, the kinetic terms are non-trivial, and thus a way to guarantee the classical limit is approached is, as we did with minisuperspace, to define the metric-dependent terms in terms of their diagonal matrix elements in a coherent-state basis, $|\pi_{cl} \gamma_{cl}\rangle$.  We see no impediment to doing this by replacing the terms in the classical Hamiltonian by operators as we did in the homnogeneous case \eqref{eq:QH-ms} of the previous section.

Thus, in the coherent state basis, the path integral for quantum gravity should read:
\begin{equation}\label{eq:GRPI}
    \langle \pi_f \gamma_f \cdots| e^{-i \hat{H} t} | \pi_i \gamma_i \cdots\rangle = \int_{\pi_i,\gamma_i}^{\pi_f,\gamma_f} \, {\cal D}\pi \, {\cal D}\gamma \, \cdots \, e^{i \int d^4x \left[\frac{1}{2}(\dot{\gamma}_{ij}\pi^{ij} - \dot{\pi}^{ij}\gamma_{ij}) - {\cal H}\left[\pi,\gamma\right]\right] + \cdots}
\end{equation}
where the various elipses represent the state quantum numbers on the left side of the equation, and the measure parts and action parts for any and all matter fields on the right side.  This path integral can be though of as a functional of the final values $\pi_f$ and $\gamma_f$, and this functional is sharply peaked around the classical expectation values of these functions. The key point is that the six degrees of freedom and their conjugates only produce (in expectation value) six pairs of Hamilton equations of motion, which correspond to the spatial parts of Einstein's equations.

Minimizing the variation of the classical action with respect to the metric led to the equations $\sqrt{-g}(G^{\mu\nu} - 8\pi G_N T^{\mu\nu}) = 0$.  In synchronous gauge, we have found only the spatial equations emerge from the quantum field theory in the classical limit.  Let's package the loosening of restriction as we did for minisuperspace:
\begin{eqnarray*}
G^{00} &=& {8\pi}G_N T^{00} + {8\pi}G_N\frac{\mathbb{H}}{\sqrt{-g}} \\
G^{0i} &=& {8\pi}G_N T^{0i} + {8\pi}G_N\frac{{\mathbb{P}}^i}{\sqrt{-g}} \\
G^{ij} &=& {8\pi}G_N T^{ij} 
\end{eqnarray*}
for, as of yet, arbitrary functions $\mathbb{H}$ and $\mathbb{P}^i$.  In this language, we define an auxiliary energy-momentum tensor
\begin{equation}
    T_{\rm aux}^{\mu\nu} = \frac{1}{\sqrt{-g}}
    \begin{pmatrix}
    \mathbb{H} & \mathbb{P}^1 & \mathbb{P}^2 & \mathbb{P}^3 \\
    \mathbb{P}^1 & 0 & 0 & 0 \\
    \mathbb{P}^2 & 0 & 0 & 0 \\
    \mathbb{P}^3 & 0 & 0 & 0 
    \end{pmatrix}
\end{equation}
 Now from these constructed classical equations, we can find  restrictions on $T_{\rm aux}$ by noting that the tensor $T$ is covariantly conserved and $G$ satisfies the Bianchi identity.  Thus, since $\nabla_\mu (G^{\mu\nu} - 8\pi G_N T^{\mu\nu} ) = 0$, then:
\begin{equation}
    0 = \nabla_\mu T_{\rm aux}^{\mu\nu} = \partial_\mu T_{\rm aux}^{\mu\nu} + \Gamma_{\mu\lambda}^{\mu} T_{\rm aux}^{\lambda\nu} + \Gamma_{\mu\lambda}^{\nu} T_{\rm aux}^{\mu\lambda}
\end{equation}
These four equations simplify in this gauge, as $\Gamma_{00}^{0} = \Gamma_{0i}^{0} = \Gamma_{00}^{i} = 0$.   We can use this condition to constrain the functions $\left[ \mathbb{H},\mathbb{P}\right]$.  Using the identity $\Gamma_{\mu\nu}^\mu = - \sqrt{-g} \partial_\nu (1/\sqrt{-g})$ and defining $t_{\rm aux}^{\mu\nu}\equiv \sqrt{-g}T_{\rm aux}^{\mu\nu}$, we can write the $\nu=0$ and $\nu=i$ equations respectively as
\begin{eqnarray*}
\partial_0 t_{\rm aux}^{00} + \partial_i t_{\rm aux}^{i0} = 0 \\
\partial_0 t_{\rm aux}^{0i} + 2 \Gamma_{j0}^{i} t_{\rm aux}^{j0} = 0
\end{eqnarray*}
which (noting that $\gamma_{ij} \equiv g_{ij}$, $\gamma = -g$, and $\gamma^{ij}\gamma_{jk}=\delta^i_k$ in this gauge) simplifies further to
\begin{eqnarray*}
\partial_0 \mathbb{H} = - \partial_i \mathbb{P}^i \\
\partial_0 \left(\gamma_{ij}\mathbb{P}^j \right) = 0
\end{eqnarray*}
We thus see that this auxiliary {\bf shadow} matter is made up of three time-independent functions and one whose time dependence is fixed by the other three (and the metric). 

\subsection{Cosmological Implications}

We will discuss some preliminary observations about the cosmological implications of this additional source term and leave a fuller analysis of the cosmology to a companion paper \cite{KMPRST} and of the effects on non-linear general relativity to future work.

To analyze the cosmological effects, we expand the metric around a homogeneous background to linear order in fields as in cosmological perturbation theory:
\begin{equation}
    ds^2 = - dt^2 + a(t)^2 (\delta_{ij} + h_{ij})dx^idx^j
\end{equation}
where we further break up $h_{ij}$ into irreducible components of the three-dimensional Euclidean group: two scalars $h$ and $\eta$, one divergenceless vector $w_i$, and one transverse-traceless tensor $s_{ij}$
\begin{equation}
    h_{ij} = h \delta_{ij} + D_{ij}\eta + (\partial_i w_j + \partial_j w_i) + s_{ij})
\end{equation}
where the differential operator $D_{ij} = \partial_i \partial_j/\nabla^2 - (1/3) \delta_{ij}$ (and $\nabla^2$ is the standard spatial Laplacian).  The inverse derivatives will be defined such that for the Fourier transforms of the fields ({\it e.g.}, $\tilde{\eta}(k)$ is the transform of $\eta(x)$), this differential operator becomes $\tilde{D}_{ij} = \hat{k}_i\hat{k}_j - \delta_{ij}/3$.

Because the different representations decouple at linear order, we can focus on the scalar perturbations.  Following \cite{Ma:1995ey} (but switching from conformal time to proper time), we can write the spatial part of the source-free Einstein equations, $G^i_{j} = 0$, in terms of the scalar modes.  In Fourier space, we define
\begin{equation}
    h_{ij}(\vec{x},t) = \int d^3 k e^{i\vec{k}\cdot\vec{x}} \left(\hat{k}_i\hat{k}_j \tilde{h}(\vec{k},t) + (\hat{k}_i\hat{k}_j - \frac{1}{3} \delta_{ij})6 \tilde{\eta}(\vec{k},t) \right)
\end{equation}
The two scalar equations are the trace part and the longitudinal traceless part of the space-space equations:
\begin{eqnarray}
    \ddot{\tilde{h}} + 3 \frac{\dot a}{a}\dot{\tilde{h}} - 2 \frac{k^2}{a^2}\tilde{\eta} = 0 \label{eq:GiiTrace}\\
    \ddot{\tilde{h}} + 3 \frac{\dot a}{a}\dot{\tilde{h}} + 6\left(\ddot{\tilde{\eta}} + 3 \frac{\dot a}{a}\dot{\tilde{\eta}} \right) - 2 \frac{k^2}{a^2}\tilde{\eta} = 0 \label{eq:GiiTransverse}
\end{eqnarray}
The first thing to note is that there are no wave-like solutions to these equations.  Subtracting \eqref{eq:GiiTrace} from \eqref{eq:GiiTransverse} produces an equation for $\tilde{\eta}$ which has two solutions: a constant and a solution in which  $\dot{\tilde{\eta}}\sim 1/a^3$.  The constant solution can be inserted in \eqref{eq:GiiTrace} and we see that during matter domination, $\tilde{h}\propto a$ is a solution.  Thus, whatever sources the expansion and the initial perturbations (whether real or shadow matter), the metric perturbations grow linearly during 'matter' domination.

We can also think of the source terms $\mathbb{H}\equiv\mathbb{H}_0 + \delta\mathbb{H}$ and $\mathbb{P}^i$ in perturbation theory, where we assume $\mathbb{H}_0$ as homogeneous and constant, and $\delta\mathbb{H}$ and $\mathbb{P}^i$ as small inhomogeneous perturbations.  While $\mathbb{H}_0$ could be the dominant homogeneous contribution to the expansion rate, we could ask about the evolution of the perturbative sources.  Assuming $\mathbb{H}_0$ is the only homogeneous contribution, the time-time equation at zeroth order is
\begin{equation}
    \left(\frac{\dot{a}}{a}\right)^2 = \frac{8\pi}{3} G_N \frac{\mathbb{H}_0}{a^3}
\end{equation}
whereas the time-time and the longitudinal part of the time-space equations of motion at linear order are
\begin{eqnarray}
    \frac{\dot{a}}{a}\dot{\tilde{h}} - 2\frac{k^2}{a^2}\tilde{\eta} &=& 8\pi G_N \frac{1}{a^3}\left[-\frac{1}{2}\tilde{h}\mathbb{H}_0 + \delta\mathbb{H}  \right] \\
   - 2 i \frac{k}{a^2}\dot{\tilde{\eta}} &=& 8\pi G_N \frac{\mathbb{P}_{\parallel}}{a^3} \,.
\end{eqnarray}
where $\delta\mathbb{H}$ and $\mathbb{P}_{\parallel}\equiv\hat{k}_i \mathbb{P}^i$ are taken to be in Fourier space.  The leading growth of $\tilde{h}$ is cancelled in the first of these equations. The contant part of $\delta\mathbb{H}$ falls off as $1/a$ relative to $\tilde{h}$, {\it i.e.}, effectively redshifts with the background density. The perturbation $\mathbb{P}_{\parallel}$ also only contributes to the dying mode of $\dot{\tilde{\eta}}$ and $\mathbb{P}_{\parallel}\sim 1/a^2$.  This is consistent with the constraint equations
\begin{eqnarray*}
\partial_0 \delta\mathbb{H} = - i k \mathbb{P}_{\parallel} \\
\partial_0 \left(a^2\mathbb{P}_{\parallel} \right) = 0
\end{eqnarray*}

\section{Discussion}
By all accounts, quantum field theory is the correct underlying theory of nature and classical physics is a limit of this quantum theory. As classical observers, we can readily observe and obtain the classical equations of motion. We are then faced with the task of obtaining the correct quantum mechanical description from the known classical equations. But, since classical physics is a limit of quantum mechanics, this is a tricky inverse problem. 

The conventional approach to this inverse problem has been to take a classical theory and replace the Poisson brackets that generate its equations of motion by various quantum commutators and use these relations to reconstruct the underlying quantum theory. While this approach works for particle mechanics and scalar field theories, we recognize that it fails for a broader class of quantum field theories. As is well known, in theories with fermions, the commutators need to be replaced with anti-commutators. In gauge theories, given gauge redundancies, some gauge degrees of freedom need to be suitably fixed in order for the theory to yield sensible results. When these ``rules'' are directly applied to general relativity, the procedure yields a trivial Hamiltonian that is manifestly incorrect. There are thus no rigid and sacred set of rules that ``derives'' a quantum theory from the known classical dynamics of the theory. Instead, the correct prescription is to see if a given quantum field theory ({\it i.e.} either a Hamiltonian or path integral description) is logically consistent and yields a classical limit that is consistent with observation. 

With this point of view, there arises an interesting possibility - since classical physics is a limit of quantum mechanics, could the quantum theory permit more freedom than inferred by the classical observer? In this paper, we have seen an example of such freedom where we see that certain Hamiltonian constraints that are imposed on the theory by the classical equations of motion do not exist at the quantum level. The principal reason for this relaxation is that the entire quantum evolution is described by just one equation - the  Schr\"{o}dinger Equation. This is a first order differential equation and thus given any initial state, it is able to describe its time evolution. This is unlike the classical theory where additional constraint equations are imposed on the system on top of the dynamical equations of motion. 

In this paper, we have shown that this relaxation permits a broader class of initial states in general relativity that result in gauge invariant time evolution. In a companion paper, we explore the consequences of such relaxation in electromagnetism. These states do not satisfy the Hamiltonian constraint of general relativity. But, their quantum mechanical evolution is such that in the classical limit of the theory, these states would behave as though they were a peculiar covariantly conserved classical stress tensor. But, there are no new degrees of freedom associated with this stress tensor - it is simply a quantum state of the metric. 

In a cosmological setting, the homogeneous piece of this classical stress tensor mimics the evolution of a homogeneous dark matter field. Interestingly, this component could have either sign. Irrespective of the sign, the associated perturbative degrees of freedom are gravitons which have positive kinetic energy terms. If this component had a negative sign, it would violate the null energy condition without any of the disastrous ghost-like instabilities often associated with such sources. Consequently, these gravitational states provide a trivial way to violate the null energy condition and yield interesting gravitational phenomena such as cosmic bounces. 

The inhomogeneous pieces of these components are phenomenologically interesting. First, in the linear regime, they grow just like dark matter and may have cosmological significance. In this paper, we have not analyzed the behavior of these inhomogeneous components in the nonlinear regime. This is however an important task since it could lead to cosmologically relevant signatures. Further, since the inhomogeneous components can also locally violate the null energy condition, they may be able to support exotic phenomena such as wormholes. 

An important feature of these initial states is that their effects can be dynamically  obliterated from our observable universe by a period of inflation. Conventional classical general relativity is thus an attractor solution if there is a period of inflation. This is unsurprising since inflation erases sensitivity to initial conditions and the nontrivial quantum effects identified by us in this paper are tied to the initial quantum state of the universe. It is thus of great interest to develop the full  phenomenology (gravitational and electromagnetic) of these quantum states since those may give rise to observational effects. If such effects are observed, they would rule out an inflationary period in the early universe, forcing us to re-explore the origins of our universe.

\section*{Acknowledgements}
We thank Peter Graham, Vivian Poulin, Tristan Smith,  Raman Sundrum and Jed Thompson for fruitful discussions. This work was supported by the U.S.~Department of Energy~(DOE), Office of Science, National Quantum Information Science Research Centers, Superconducting Quantum Materials and Systems Center~(SQMS) under Contract No.~DE-AC02-07CH11359. D.E.K.\ and S.R.\ are supported in part by the U.S.~National Science Foundation~(NSF) under Grant No.~PHY-1818899.
S.R.\ is also supported by the~DOE under a QuantISED grant for MAGIS.
The work of S.R.\  was also supported by the Simons Investigator Award No.~827042. T.M.\ is supported by the World Premier International Research Center Initiative (WPI) MEXT, Japan, and by JSPS KAKENHI grants JP19H05810, JP20H01896,  JP20H00153, and JP22K18712. 

\bibliographystyle{unsrt}

\bibliography{references}

\begin{appendices}

\section{ Quantum Field Theory on a Manifold}\label{sec:Mani}

Let's clarify the mathematical structure we are assuming for the quantum theory of gravity.  We want a quantum field theory of a metric tensor coupled to other quantum fields and living on a fixed manifold.  The manifold is taken to be a four-dimensional manifold with topology $\mathbb{R}\times\Sigma$. A conventional choice for $\Sigma$ is $\mathcal{R}^3$ but other choices are possible.   There is one question the field theory should be able to answer: given an initial quantum state on $\Sigma$, how does this state evolve in ``time''?

The subtlety in answering this question is that the manifold does not have a preferred time coordinate, or potentially even slicing, and physical questions should have answers which are time reparameterization invariant.  In addition, of the ten fields in the metric tensor, only two are dynamical, whereas four are pure gauge degrees of freedom, and the other four are fixed by classical constraints.  All of this extra degeneracy should be taken care of in the quantum theory without generating additional degrees of freedom while maintaining the reparameterization invariance of the manifold.  We will describe the theory using the path integral as it offers an easier path to handle these issues.

To define the path integral, we begin by taking the manifold and picking some arbitrary foliation $f\left(t\right)$ where the function $f\left(t\right)$ is a one to one function that assigns to the parameter $t$ a set of points $\Sigma\left(t\right)$ whose topology is $\Sigma$. These foliations are the ``time'' slices that we have picked for this manifold. On each $\Sigma\left(t\right)$, pick some spatial coordinates $\left(x^{1}, x^{2}, x^{3}\right)$. The above assignments are performed at the level of the manifold and these do not require specification of the metric. On this foliation, the quantum field that is the metric tensor is expressed in the form: 
\begin{equation}
    g_{\mu \nu} \rightarrow ds^2 = - N^2 dt^2 + N_{i} dt dx^{i} + s_{ij} dx^{i} dx^{j}
\end{equation}
Similarly, the non-gravitational quantum fields in the theory are described as various vector, scalar or spinor fields on this foliation. Given these, we can define the usual Einstein-Hilbert Lagrangian $\mathcal{L}_{GR}$ and matter Lagrangian for the theory. Note that the definition of the Lagrangian requires a foliation since the Lagrangian is implicitly a function of space-time and contains space-time derivatives.  

Now comes the physics via the path integral.  For simplicity, we will focus on the pure gravitational part of the theory. From the Lagrangian, we know that the fields $N$ and $N_i$ do not contain conjugate momenta. These degrees of freedom are not physical and they are undetermined. Thus, one can arbitrarily set, for example, $N=1$ and $N_{i} = 0$ and assert that the matrix elements of the time evolution operator of the degrees of freedom  $s_{ij}$ from time $t_1$ to $t_2$ are given by the path integral: 

\begin{equation}
    \langle s^{f}_{ij} | T\left(t_2; t_1\right) | s^{i}_{ij} \rangle = \int_{s_{ij}\left(t_1\right) = s^{i}_{ij}}^{s_{ij}\left(t_2\right) = s^{f}_{ij}} D s_{ij} \, e^{i S} 
    \label{eq:genPI}
\end{equation}
where the states $| s^{i,f}_{ij}\rangle$ are field basis states\footnote{A similar construction can also be employed for coherent states. We will in fact adopt this construction from Section \ref{sec:mini} onwards.} and $S = \int_{t_1}^{t_2} d^4 x \, \mathcal{L}_{GR}$ is the Einstein-Hilbert action. Notice that this path integral does not involve integration over the functions $N$ and $N_i$ which have been fixed. The integral is over the six degrees of freedom $s_{ij}$. The above path integral is well defined and it can be calculated. 

The next task is to show that this construction where we took an arbitrary foliation and made choices of the functions $N, N_i$ gives rise to physics that is invariant under these choices. To see this, we first fix the  initial and final slices $\Sigma\left(t_1\right)$ and $\Sigma\left(t_2\right)$, respectively. Now pick any arbitrary foliation in the middle.  This amounts to keeping the parameters $t_1$ and $t_2$ fixed on the manifold  while reparameterizing $t$ in between $t_1$ and $t_2$. The transformation from the initial foliation to this new foliation is a coordinate transformation under which all the fields have well defined  transformations. These transformations will in  general mix the elements of the metric  to yield a new metric $\tilde{g}_{\mu \nu} =  \frac{\partial x^{\alpha}}{\partial x^{'\mu}}\frac{\partial x^{\beta}}{\partial x^{'\nu}} g_{\alpha \beta}$. Thus, the metric components $\tilde{g}_{00}, \tilde{g}_{0i}$ will be different from the initial values ($-1$ and $0$, respectively) that we had arbitrarily chosen. We thus recognize that coordinate transformations allow us to change between different values of $N, N_i$. Fixing $N, N_i$ is thus a gauge fixing procedure where we pick the coordinates on which the path integral is performed - without fixing the coordinates, one cannot evaluate the path integral since the integral is a function known in terms of coordinates.

The change in time slices also means that fields that were defined to be at the ``same time'' in one foliation will be mapped to unequal points in ``time'' in other foliations. To keep track of these changes, one can discretize the spatial manifold $\Sigma$ and track the fields values at discrete spatial points. Once this is performed, the path integral is now an integral over functions defined at these discrete spatial points with a different parameterization of time when the foliation is changed. The metric that appears in the path integral  \eqref{eq:genPI} will contain more non-trivial elements.  However, the number of physical functions that are allowed to vary in the path integral will still be six. The measure of the path integral does not change under the reparameterization since we are still integrating over the same field values at these spatial points - all that has been done is a relabelling of the time parameter. Since the action of the theory is  invariant under this reparameterization, the transition matrix elements  $\langle s^{f}_{ij} | T\left(t_2; t_1\right) | s^{i}_{ij} \rangle$ are invariant.

Second, let us return to the original foliation that we had picked. Retaining $N_i = 0$, pick $N$ to be any arbitrary function of $t$. Due to the reparameterization invariance of the theory, one can show that $\langle s^{f}_{ij} | T\left(t_2; t_1\right) | s^{i}_{ij} \rangle =  \langle s^{f}_{ij} | T\left(\tilde{t}_2; \tilde{t}_1\right) | s^{i}_{ij} \rangle$ where the time labels $\tilde{t}$ are obtained by solving the equation $d\tilde{t} = N\left(t\right) dt$. We thus understand the meaning of $N$ - it simply reflects a choice of time parameterization. Even though \eqref{eq:genPI}  is explicitly dependent on the choice of time parameterization, the physics that it describes, namely, the relative phase between various quantum states as described by the transition matrix elements, is invariant as long as we recognize that the time evolution operator  \eqref{eq:genPI} is suitably changed when the time parameter is changed by a new foliation. The physics described by a parameterization $t$ and a function $N\left(t\right)$ is identical to the physics described by the parameterization $\tilde{t}$ and the function $\tilde{N}\left(\tilde{t}\right)$ where $N\left(t\right) dt = \tilde{N}\left(\tilde{t}\right) d\tilde{t}$. Neither $N$ or $t$ are individually physical - but this differential combination is physical. Note that all other changes to the foliations and the functions $N, N_i$ in the middle of these two time slices can be constructed as a composition of these two cases and thus lead to gauge invariant physics. 

Finally, let us consider a foliation which also changes the initial and final slices $\Sigma\left(t_{1,2}\right) \rightarrow \Sigma\left(\tilde{t}_{1,2}\right)$. Since the initial and final slices now correspond to different points on the manifold, there is no canonical identification that permits us to show that the physics is invariant. This is a particularly important aspect of quantum mechanics in comparison to classical physics. In classical field theory, one could evaluate the values of various functions at the same point on the manifold and potentially compare those values using well defined coordinate transformation properties. However, a quantum state is inherently non-local since it naturally includes entanglement with modes of the field that could be localized at different points on the slices $\Sigma$ and thus there is no simple local comparison that one can do. Formally, the Hilbert space is different on different time slicing.  However, it is clear that the physics is invariant under such a foliation change since one may regard this kind of foliation change as being naturally embedded in the ``middle'' of a larger set of foliations describing the evolution of a quantum state from $\Sigma\left(t^{-}_{1}\right)$ to $\Sigma\left(t^{+}_{2}\right)$ where $t^{-}_{1} \leq t_1 \leq t \leq t_2 \leq t^{+}_{2}$. This foliation change is then already described by the situations above. 

An extremely important aspect of these coordinate transformations is that these transformations are simply functions that map one set of coordinates on the manifold to another. They are thus independent of the  basis state whose time evolution is being described by the path integral. This implies that certain metric redefinitions, which one might describe as ``gauge'' choices,  performed in classical general relativity have to be properly interpreted in the quantum context. See the discussion of this point in  \cite{Burns:2022fzs}. 

The above construction treats $\Sigma\left(t\right)$ as a time slice in performing the evolution of the system. In general relativity, given a set of time slices, for some initial data, the time evolution could result in the time slices becoming spatial. This is a fact even in classical general relativity and it cannot be avoided in the quantum theory. This implies that in certain situations the computation of the path integral for certain foliations can become ill defined since functions like the Lagrangian will contain divergent or ill defined terms as the time slice evolves into a spatial slice. But, the solution for these issues is similar to the approach taken in classical general relativity.  In the case where these issues arise due to coordinate singularities, one may choose a different time slice and perform the evaluation and see if these coordinate singularities can be avoided.

\section{Coherent States and the Path Integral}
\label{sec:Coherent}

First we construct a coherent state basis based on an arbitrary harmonic oscillator.  Independent of the width of the harmonic oscillator, we know this is an over-complete basis.  For the field operator $\hat{a}$ and its conjugate $\hat{\pi}$ with standard commutators, we define coherent states $|\pi_{cl} a_{cl} \rangle$ such that
\begin{eqnarray}
    \langle \pi_{cl} a_{cl} | \hat{a} | \pi_{cl} a_{cl} \rangle &=& a_{cl} \nonumber\\
    \langle \pi_{cl} a_{cl} | \hat{\pi} | \pi_{cl} a_{cl} \rangle &=& \pi_{cl} \label{eq:Co-vevs}
\end{eqnarray}
Now define $\hat{A} \equiv (\sigma \hat{a} + (1/\sigma) \hat{\pi})/\sqrt{2}$ and a state $|0\rangle_\sigma$ such that $\hat{A}|0\rangle_\sigma = 0$, where $\sigma$ is an arbitrary real parameter.  We know that any state
\begin{equation} \label{eq:Co-state}
    |\alpha\rangle \equiv e^{-|\alpha|^2/2} e^{\alpha \hat{A}^\dagger} |0\rangle_\sigma 
\end{equation}
(with $\alpha$ a complex number) satisfies $\hat{A}|\alpha\rangle = \alpha |\alpha\rangle$ and $\langle\alpha\rangle = 1$.  Using $\left[ \hat{A},\hat{A}^\dagger\right] =1$, and \eqref{eq:Co-vevs}, that $|\pi_{cl} a_{cl}\rangle = |\alpha\rangle$ with the real and imaginary parts of $\alpha$ defined as $\alpha^R = \sigma a_{cl}/\sqrt{2}$ and $\alpha^I = \pi_{cl}/\sqrt{2}\sigma$, respectively.

Now define the Hamiltonian: 
\begin{equation}\label{eq:QH-ms XXX}
    \hat{H} = - \frac{1}{24 M_{pl}^2} \widehat{\left[ \frac{\pi^2}{a}\right]} + \frac{1}{2} \hat{\pi}_\phi^2 \widehat{\left[\frac{1}{a^3}\right]} + \widehat{\left[a^3\right]} V(\hat{\phi})
\end{equation}
where the operators in square brackets are defined by their diagonal matrix elements in the $|\pi_{cl} a_{cl}\rangle$ basis, {\it e.g.},
\begin{equation}
    \langle \pi_{cl} a_{cl} | \widehat{\left[\frac{\pi^2}{a}\right]} | \pi_{cl} a_{cl} \rangle = \frac{\pi_{cl}^2}{a_{cl}}
\end{equation}
The matter coupling requires a different basis as for the coherent states defined around this origin, this operator is ill-defined.  For simplicity, we will stick to the pure gravity case.

Now we are poised to build the path integral using the coherent-state basis. Using the completeness relation for coherent states:
\begin{equation}
   \frac{1}{\pi} \int \left[ d\pi\right] \left[ da \right] | \pi a \rangle \langle \pi a | = 1
\end{equation}
one can break a transition matrix element into infinitesimal time steps $\delta t$ by inserting a complete set of coherent states:
\begin{equation}\label{eq:pre-PI}
    \langle \pi' a' | e^{-i \hat{H} t} | \pi a \rangle = \prod_{k=0}^N \int \left[\frac{d\pi_k}{\pi}\right]\left[da_k\right] \langle\pi_k a_k| \hat{1} - i \hat{H}\delta t |\pi_{k-1} a_{k-1} \rangle \delta(\pi_N - \pi')\delta(a_N - a')\delta(\pi_0 - \pi)\delta(a_0 - a)
\end{equation}

Let's compute one of the infinitesimal matrix elements in \eqref{eq:pre-PI}, $ \langle \pi_k a_k | \hat{1} - i\delta t \hat{H} |\pi_{k-1} a_{k-1}\rangle $.  The identity matrix element can be approximated:
\begin{equation}
    \langle\pi_k a_k | \pi_{k-1} a_{k-1}\rangle \simeq \langle\pi_k a_k|\pi_k a_k\rangle - \langle\pi_k a_k | \frac{|\pi_k a_k \rangle - |\pi_{k-1} a_{k-1}\rangle}{\delta t}\delta t
\end{equation}
Approximating the last term as a total time derivative in the vanishing step-size limit, $\sim \frac{d}{dt} \langle \pi a | \pi a \rangle (t) \delta t$, we can rewrite the full term as
\begin{multline}
    \langle \pi_k a_k | \hat{1} - i\delta t \hat{H} |\pi_{k-1} a_{k-1}\rangle \\ \simeq 1 - \left[\dot{\pi}_k \frac{\partial}{\partial \pi_{k-1}}\langle\pi_k a_k | \pi_{k-1} a_{k-1}\rangle  + \dot{a}_k \frac{\partial}{\partial a_{k-1}}\langle\pi_k a_k | \pi_{k-1} a_{k-1}\rangle \right]_{k-1 \rightarrow k} \delta t - i \langle \pi_k a_k | \hat{H} | \pi_k a_k \rangle  \delta t
\end{multline}
we can evaluate the field derivatives using the explicit form of the coherent states \eqref{eq:Co-state}, {\it e.g.},
\begin{eqnarray}
    \frac{\partial}{\partial a_{cl}} |\pi_{cl} a_{cl} \rangle = \frac{\sigma}{\sqrt{2}}\frac{\partial}{\partial \alpha^R}|\pi_{cl} a_{cl}\rangle = \frac{\sigma}{\sqrt{2}} (-\alpha^R + \hat{A}^\dagger ) |\pi_{cl} a_{cl} \rangle \\
    \frac{\partial}{\partial \tilde{a}_{cl}} \langle \pi_{cl} a_{cl} | {\pi}_{cl} \tilde{a}_{cl} \rangle\bigg\vert_{\tilde{a} \rightarrow a}  = \frac{\sigma}{\sqrt{2}}(-\alpha^R + (\alpha^R -i\alpha^I ) = -i \frac{\sigma}{\sqrt{2}} \alpha^I = -\frac{i}{2} \pi_{cl}
\end{eqnarray}
and similarly $\partial \langle \pi a | \tilde{\pi} a\rangle/\partial \tilde{\pi} \rightarrow i a/2$. Thus, for an infinitesimal time step, we have (exponentiating):
\begin{equation}
    \langle \pi_k a_k | \hat{1} - i\delta t \hat{H} |\pi_{k-1} a_{k-1}\rangle \simeq e^{ i\left(\frac{1}{2} ( \dot{a}_k \pi_k -\dot{\pi}_k a_k) - H(\pi_k,a_k) \right)\delta t }
\end{equation}
from which we can derive the path integral \eqref{eq:msPI}.

\subsection{Coherent State Basis Operators}
\label{sec: CoStOps}

Here we show how the relationship between the number-operator basis (the eigenstates of a harmonic oscillator) and the coherent state basis of operators.  We will also show that an operator in the coherent state basis is completely defined by its diagonal elements.  As we will see, one can derive all off-diagonal matrix elements from diagonal ones.  We will follow the nice notes of \cite{Wheeler:2012xx}, but see also \cite{Klauder:1994xx}.  For generalization to fields, see, for example, the classic paper by Sudarshan \cite{Sudarshan:1963ts}

Coherent states, states that are eigenvectors of the lowering operator $\hat{A}$, can be written in the number-operator basis:
\begin{equation}
    |\alpha\rangle = e^{-|\alpha|^2/2} \sum_n \frac{\alpha^n}{\sqrt{n!}}|n\rangle
\end{equation}
Define an operator $\hat{O}$ in the number-operator basis as $O_{mn} \equiv \langle m| \hat{O} |n \rangle$. And define
\begin{equation}\label{eq:coherentmatrix}
    {\cal O}_{\bar{\alpha}\beta} \equiv e^{+(|\alpha|^2 + |\beta|^2)/2} \langle \alpha | \hat{O} | \beta\rangle = \sum_{mn} O_{mn} \frac{\bar{\alpha}^m \beta^n}{\sqrt{m! n!}}
\end{equation}
writing $\alpha^*\rightarrow\bar{\alpha}$ for convenience.  One can immediately see that all of the matrix elements of $\hat{O}$ in the number-operator basis can be derived from the diagonal matrix elements in the coherent state basis:
\begin{equation}\label{eq:off-diag}
    O_{mn} = \frac{1}{\sqrt{m! n!}} \frac{\partial^m}{\partial\bar{\alpha}^m}\frac{\partial^n}{\partial\alpha^n} {\cal O}_{\bar{\alpha}\alpha} \bigg\vert_{\bar{\alpha}=\alpha = 0}
\end{equation}
One can also derive the off-diagonal matrix elements in the coherent state basis as well by plugging \eqref{eq:off-diag} into \eqref{eq:coherentmatrix}.

\end{appendices}

\end{document}